\documentclass[
aps,%
12pt,%
final,%
notitlepage,%
oneside,%
onecolumn,%
nobibnotes,%
nofootinbib,%
superscriptaddress,%
showpacs,%
centertags]%
{revtex4}

\usepackage{epsfig}
\usepackage{graphics}

\begin{document}
\title{Charmed quark fragmentation in $B$-mesons decays}

\author{\firstname{Alexey}~\surname{Novoselov}}
\email{Alexey.Novoselov@cern.ch}
\affiliation{Moscow Institute of Physics and Technology, Dolgoprudny, Russia}
\affiliation{Institute for High Energy Physics, Protvino, Russia}

\begin{abstract}
It is well known that large nonfragmentational contributions to inclusive
production of charmed particles appear at low energies.
In the case of charm production in $B$-meson decays these contributions
arise from the participation of the light valent quark from the $B$-meson
and can be easily described phenomenologically.
These contributions affect $D$-meson spectra and essentially
change ratios between yields of different $D$-meson states. 
\end{abstract}


\maketitle

Recent study of $D^{(*)}$-mesons a $\Lambda_C$-baryons production in the $e^+e^-$-annihilation
have shown that differential production cross-sections of these particles
in $10$-$90$ GeV energy range can be obtained using the framework of fragmentation functions
\cite{Novoselov:2010zz}.
One and the same nonperturbative fragmentation function was used in the whole energy range
while the energy dependence was incorporated in the perturbative fragmentation function.
Thus the nonperturbative fragmentation function depended only on the type of final particle.

The usage of the fragmentational approach for $D^{(*)}$-mesons and $\Lambda_C$-baryons production
in the $e^+ e^- \to \Upsilon(4S) \to B \bar{B} \to D^* + X, \Lambda_C + X$ process is arguable due to
the low mass of the $B$-meson. Nonetheless good agreement with the experimental data has been attained
for the $\Lambda_C$ momentum distribution in this process. This is not the case for the
$D^{(*)}$-mesons, whose spectrum has an additional contribution with the momentum higher than predicted.

One of the possible explanations is that the light valent quark in the $B$-meson can take part
in the charmed meson formation and lead to the higher momentum of the latter than the fragmentational
mechanism. This valent quark can not affect the baryon production as an ($ud$)-diquark-spectator is
needed for the $\Lambda_C$ formation. This mechanism must lead to the correlation between quark compositions
of the decaying and formating mesons.

The BABAR collaboration have measured inclusive $B^-$ and $\bar{B}^0$ decays
to flavor-tagged $D$-mesons \cite{Aubert:2006mp}.
Momentum distributions for $D^+$, $D^-$, $D^0$ and $\bar{D}^0$ in the $B$ rest frame
are presented. These data allow to check the aforesaid conjecture.
Though experimental data concern scalar mesons which appear in the decays of vector $D^*$-mesons
more often than by themselves. That is why it is important to account for the
$D^*$-mesons decays.

\begin{figure}
  \includegraphics[width=3in]{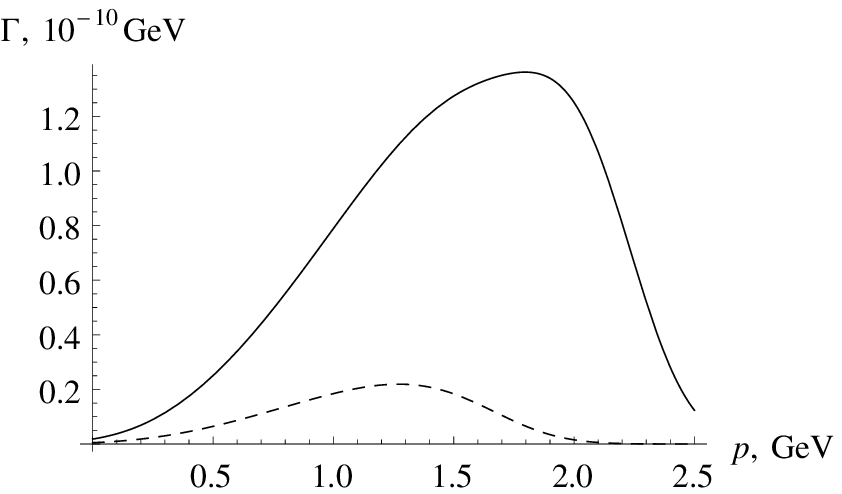}\\
  \caption{.}
  \label{fig:el-w}
\end{figure}

Let us first consider the $D^*$ production. The same will refer to the scalar $D$-mesons
born directly, not in the $D^*$ decay.
To fix the idea we will talk about the $B^-$-meson decays. $\bar{B}^0$ decays are treated
similarly with the substitution of $\bar{u}$-quark with $\bar{d}$.
$B^-$ decays through weak channel and we are interested in the $c$ and $\bar{c}$ momentum
spectra in $b \to c(p_c)\, f_1\, f_2$ and $b \to c\, \bar{c}(p_{\bar{c}})\, s$ processes.
Differential widths of this processes are equal
$\Gamma(p_c,m_1,m_2)$ and $\Gamma(p_{\bar{c}},m_c,m_s)$ while
\begin{eqnarray}
  \frac{d\Gamma}{dp}(p,m_1,m_2) &=& \frac{G_F^2 p^2}{12 \pi ^3 E\left(m_b^2+m_c^2-2 m_b E\right){}^3}\theta \left(m_b^2-2 m_b E+m_c^2-\left(m_1+m_2\right){}^2\right)
  \nonumber
   \\&&
   \sqrt{m_b^2+m_c^2-2 m_b E-\left(m_1-m_2\right){}^2} \sqrt{m_b^2+m_c^2-2 m_b E-\left(m_1+m_2\right){}^2}
  \nonumber
   \\&&
   \left[3 m_b^6 E-2 m_b^5 \left(9 m_c^2+8 p^2\right)+m_b^4 E\left(45 m_c^2+28 p^2\right)+\right.
   \label{eq:G1}
   \\&&
   + m_b^2 E\left(-4 p^2 \left(-7 m_c^2+m_1^2+m_2^2\right) + 45 m_c^4-3 \left(m_1^2-m_2^2\right){}^2\right)
   \nonumber
  \\&&
   +2 m_b^3 \left(-36 p^2 m_c^2-30 m_c^4+p^2 \left(m_1^2+m_2^2-8 p^2\right)\right) +3 m_c^2 E \left(m_c^4-\left(m_1^2-m_2^2\right){}^2\right)
  \nonumber
   \\&& \left.
   +2 m_b \left(p^2 \left(-8 m_c^4+\left(m_1^2+m_2^2\right) m_c^2+\left(m_1^2-m_2^2\right){}^2\right)-9 m_c^6+3 \left(m_1^2-m_2^2\right){}^2 m_c^2\right) \right],
  \nonumber
\end{eqnarray}
where $E=\sqrt{(p^2+m_c^2)}$, and $m_1$ and $m_2$ denote masses of particles $f_1$ and $f_2$.
These particles may be pair of light quarks $\bar{u}$ and $d$, $\bar{c}s$-pair or leptonic
pairs $e \bar{\nu}_e$, $\mu \bar{\nu}_{\mu}$ and $\tau \bar{\nu}_{\tau}$.
Masses of these particles are assumed to be
$m_b=5.0$ GeV, $m_с=1.7$ GeV, $m_s=m_e=m_{\mu}=0$, $m_\tau=1.8$ GeV.
For the quark sector the corresponding widths are enhanced by the color factor of $3$.
Momentum distribution of the $\bar{c}$-quark obviously coincide with those of
the $c$-quark accompanied by the $\bar{c} s$ pair production.
Momentum of the $b$-quark within $B$-meson is supposed to be Gaussian with
the width of $400$ MeV. Resulting spectra of the $c$ and $\bar{c}$-quarks
are presented in Fig. \ref{fig:el-w}.

For the momenta distribution of charmed mesons the factorizational relation is used:
\begin{eqnarray}
  \frac{d\sigma_D}{dx} &=& \frac{d\sigma_{c (\bar{c})}}{dx} \otimes D(z),
\end{eqnarray}
where $x = p_D/p_c^{max}$ is the reduced momentum of the meson and $D(z)$ is the fragmentational
function. For both $D$ and $D^*$-mesons the KLP fragmentation function \cite{Kartvelishvili:1977pi}
is used to describe the fragmentational contribution.
The only parameter of this function $\alpha=3.6$ was determined by fitting the high energy data on
$D^{(*)}$-meson production in $e^+e^-$-annihilation. While it is important to account for
perturbative part of fragmentational function at high energies, it can be neglected in our case
as it does not appreciably differ from the $\delta$-function at the $m_B$-scale.

For the $D^{(*)0}$ production we modify the fragmentation function in the following way:
\begin{eqnarray}
  D^{np}_{D^0}(z) &=& {1 \over 2} (1-A) \frac{\Gamma (\alpha +3)}{\Gamma (\alpha +1)} z^{\alpha } (1-z) + A \delta(1-z),
  \label{eq:npFF0}
\end{eqnarray}
where first term is traditional KLP function and the second corresponds to the
recombination with the spectator. For the $D^{(*)+}$-mesons recombination is not possible as
there is no $\bar{d}$-quark in the $B^-$. Thus the KLP term is the only term in the
$D^{(*)+}$ fragmentation function:
\begin{eqnarray}
  D^{np}_{D^+}(z) &=& {1 \over 2} (1-A) \frac{\Gamma (\alpha +3)}{\Gamma (\alpha +1)} z^{\alpha } (1-z).
  \label{eq:npFFplus}
\end{eqnarray}

One can see that such modification changes not only the shape but also the
normalization of the fragmentational function. Nevertheless the sum of the
fragmentation functions of $D^{(*)+}$ and $D^{(*)0}$ does not change as we
still have one $c$-quark to hadronize.
The $D^{(*)-}$ and $\bar{D}^{(*)0}$ formation is possible throw the
$\bar{c}$-quark fragmentation. The recombination is not possible for them
and the same functions with $A=0$ are used.

Charmed quark hadronizes to $D^*$ with larger probability than to $D$. The ratio of these
probabilities is denoted by $C$ and was varied to describe the experimental data
most accurately. $D^{*} \to D + \pi$ and $D^{*} \to D + \gamma$ decays occur with
well known branching fractions and are elaborated in \cite{Cacciari:2005uk}.

Regarding the aforementioned points partial widths of the decays considered are
written down as follows:
\begin{eqnarray}
  \nonumber
  Br_{B^- \to D^0 + X} &=& \frac {\Gamma^{tot}_{B \to D + X}} {\Gamma_C + \Gamma_{\bar C}} {1\over2}\left((1+A)(1+C Br_{D^{*0} \to D^0})+(1-A) C Br_{D^{*+} \to D^0})\right)
  \\
  \nonumber
  Br_{B^- \to D^+ + X} &=& \frac {\Gamma^{tot}_{B \to D + X}} {\Gamma_C + \Gamma_{\bar C}} {1\over2}\left((1-A)(1+C Br_{D^{*+} \to D^+})\right)
  \\
  \nonumber
  Br_{B^- \to \bar D^0 + X} &=& Br_{\bar B^0 \to \bar D^0} = \frac {\Gamma^{tot}_{B \to D + X}} {\Gamma_C + \Gamma_{\bar C}} \frac {\Gamma_{\bar C}} {\Gamma_C} {1\over2}\left((1+C (Br_{\bar D^{*0} \to \bar D^0} + Br_{D^{*-} \to \bar D^0}))\right)
  \\
  \nonumber
  Br_{B^- \to D^- + X} &=& Br_{\bar B^0 \to D^-} = \frac{\Gamma^{tot}_{B \to D + X}} {\Gamma_C + \Gamma_{\bar C}} \frac {\Gamma_{\bar C}} {\Gamma_C} {1\over2}\left((1+C Br_{D^{*-} \to \bar D^-})\right)
  \\
  \nonumber
  Br_{\bar B^0 \to \bar D^0 + X} &=& \frac{\Gamma^{tot}_{B \to D + X}} {\Gamma_C + \Gamma_{\bar C}} {1\over2}\left((1+C (Br_{\bar D^{*0} \to \bar D^0} + Br_{D^{*-} \to \bar D^0}))\right)
  \\
  Br_{\bar B^0 \to D^+ + X} &=& \frac{\Gamma^{tot}_{B \to D + X}} {\Gamma_C + \Gamma_{\bar C}} {1\over2}\left((1+A)(1+C Br_{D^{*+} \to D^+})\right),
\end{eqnarray}
where $\Gamma^{tot}_{B \to D + X}$ it the total width of the $B$ to all $D$-states decay.

Best agreement with the experimental distributions is found with the
values $A=0.6$ and $C=2$.
This value of $C$ is in between the naive spin counting estimation ($C=3$) and the value
$C=1.4$, measured in the $Z$-boson decays \cite{Barate:1999bg}.
The comparison of the partial widths with the experimental results is presented in Tab. \ref{table:t1}.
All the values are within the error ranges. Many experimental uncertainties cancel
when considering ratios of yields of charm-correlated and anticorrelated particles of every type.
This ratios are shown in Tab. \ref{table:t2} and also do not contradict the experimental values.

\begin{table*}
\begin{center}
\caption{Partial widths of the $B \to D X$ decays.
Experimental values are taken from \cite{Aubert:2006mp}.}
\label{table:t1}
\begin{tabular}{|c|c|c|}
  \hline
  $Br_{B^- \to D^0 + X }$ & 77.0 & $78.6\pm1.6\pm2.7^{+2.0}_{-1.9}$ \\
  $Br_{B^- \to D^+ + X }$ & 9.4 & $9.9 \pm 0.8 \pm 0.5^{+0.8}_{-0.7}$ \\
  $Br_{B^- \to \bar D^0 + X }$ & 7.8 & $8.6\pm0.6\pm 0.3^{+0.2}_{-0.2}$ \\
  $Br_{B^- \to D^- + X }$ & 2.9 & $2.5 \pm 0.5 \pm 0.1^{+0.2}_{-0.2}$ \\
  $Br_{\bar B^0 \to D^0 + X }$ & 48.6 & $47.4\pm2.0\pm1.5^{+1.3}_{-1.2}$ \\
  $Br_{\bar B^0 \to D^+ + X }$ & 37.8 & $36.9 \pm 1.6 \pm 1.4^{+2.6}_{-2.3}$ \\
  $Br_{\bar B^0 \to \bar D^0 + X }$ & 7.8 & $8.1\pm1.4\pm0.5^{+0.2}_{-0.2}$ \\
  $Br_{\bar B^0 \to D^- + X }$ & 2.9 & $2.3 \pm 1.1 \pm 0.3^{+0.2}_{-0.1}$ \\
  \hline
\end{tabular}
\end{center}
\end{table*}

\begin{table*}
\begin{center}
\caption{The fraction of anti-correlated $B$-meson decays compared to the
BABAR collaboration data \cite{Aubert:2006mp}.}
\label{table:t2}
\begin{tabular}{|c|c|c|}
  \hline
  $Br_{B^- \to \bar D^0 + X } / Br_{B^- \to D^0 + X }$ & 0.092 & $0.098\pm0.007\pm0.001$ \\
  $Br_{B^- \to D^- + X } / Br_{B^- \to D^+ + X }$ & 0.237 & $0.204\pm0.035\pm0.001$ \\
  $Br_{\bar B^0 \to \bar D^0 + X } / Br_{\bar B^0 \to D^0 + X }$ & 0.138 & $0.146\pm0.022\pm0.006$ \\
  $Br_{\bar B^0 \to D^- + X } / Br_{\bar B^0 \to D^+ + X }$ & 0.072 & $0.058\pm0.028\pm0.006$ \\
  \hline
\end{tabular}
\end{center}
\end{table*}

\begin{figure}[!ht]
    \begin{center}
    \includegraphics[width=4in]{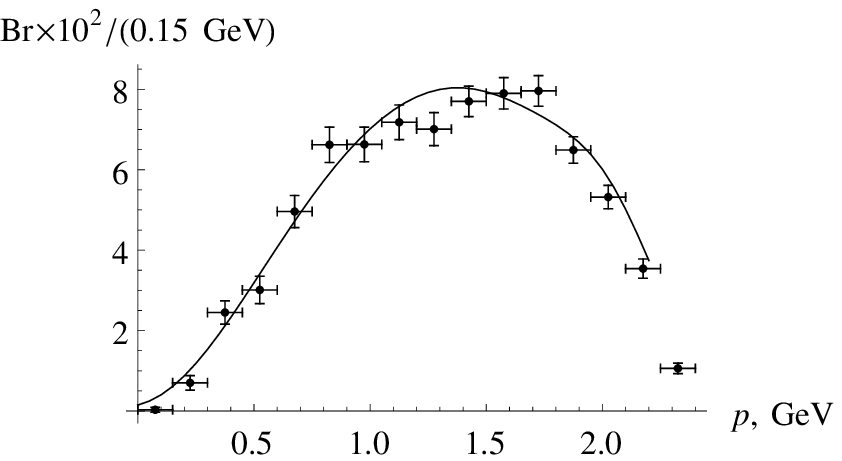}
    \includegraphics[width=4in]{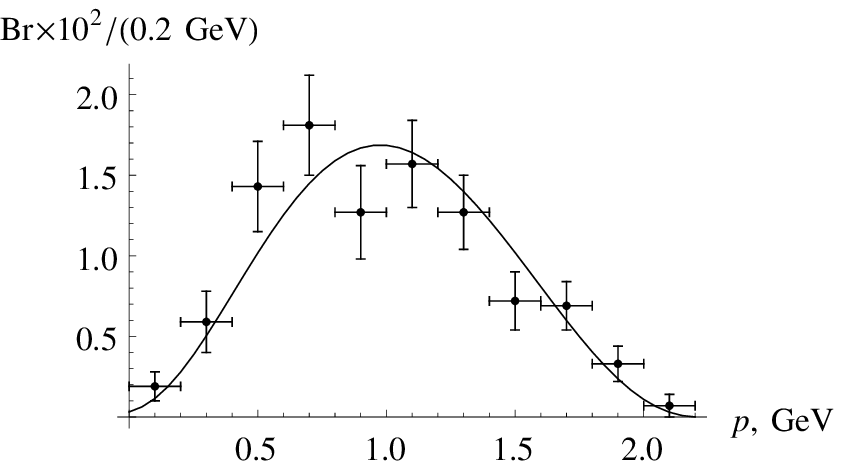}
    \includegraphics[width=4in]{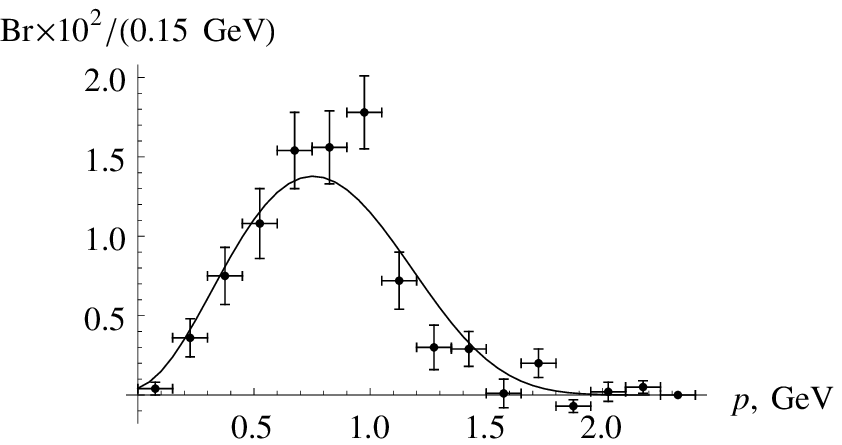}
    \includegraphics[width=4in]{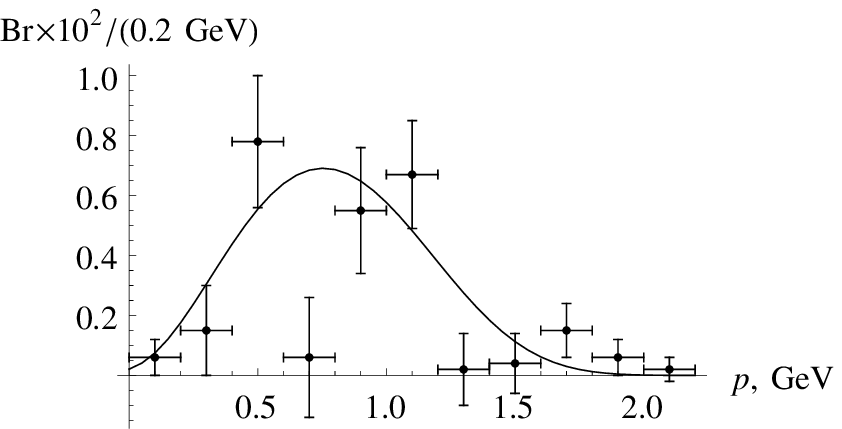}
   \caption{Momentum distributions of the $B^- \to D^0$, $B^- \to D^+$, $B^- \to \bar D^0$ and $B^- \to D^-$ decays (from top to bottom).}
   \label{fig:BmD}
    \end{center}
\end{figure}

\begin{figure}[!ht]
    \begin{center}
    \includegraphics[width=4in]{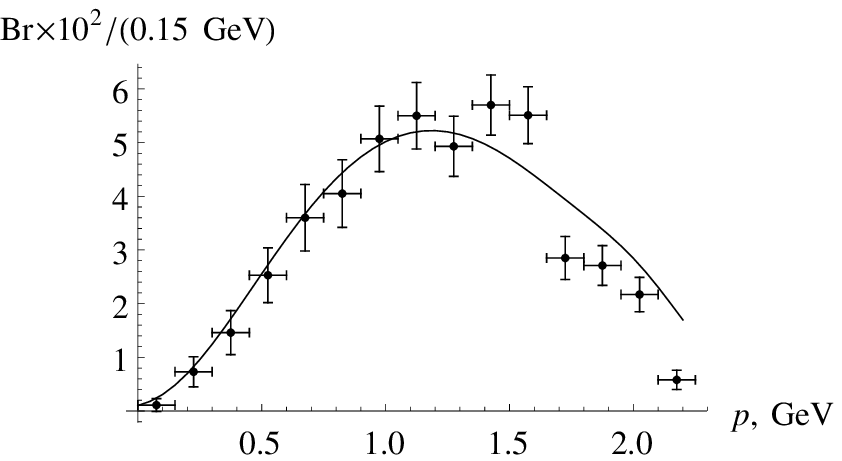}
    \includegraphics[width=4in]{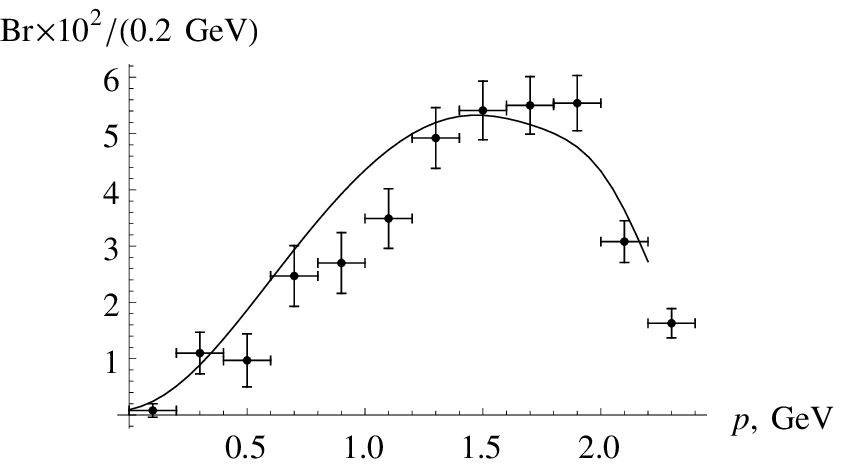}
    \includegraphics[width=4in]{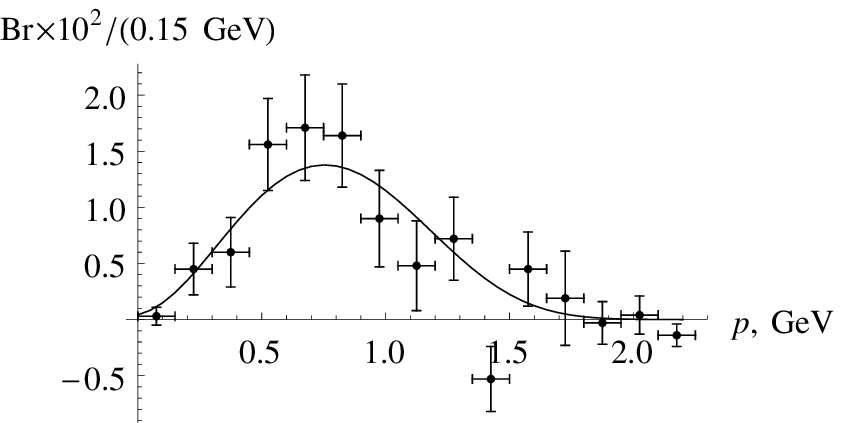}
    \includegraphics[width=4in]{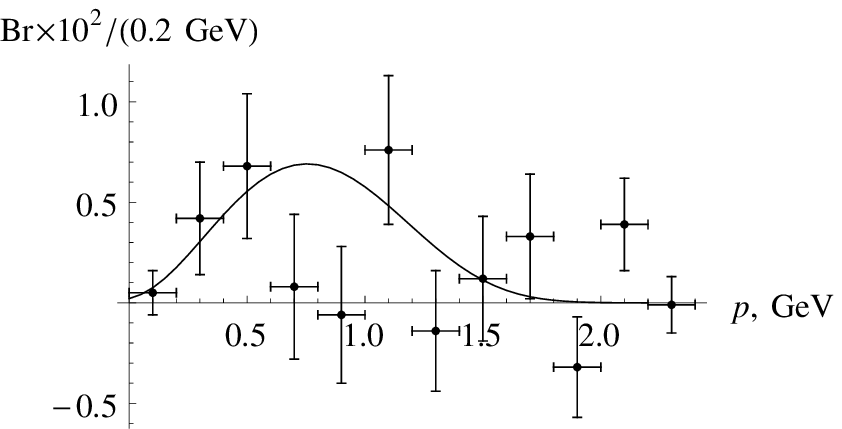}
    \caption{Momentum distributions of the $\bar B^0 \to D^0$, $\bar B^0 \to D^+$, $\bar B^0 \to \bar D^0$, $\bar B^0 \to D^-$ decays (from top to bottom).}
    \label{fig:B0bD}
    \end{center}
\end{figure}

Apart from the partial widths momentum distributions of the $D$-mesons in the $B$ rest frame
can be obtained. They are presented in Fig. \ref{fig:BmD} and \ref{fig:B0bD}.

To sum it up, it was shown that the fragmentational approach has to be supplemented
with the recombination terms in $B$-meson decays.
The part of recombination is significant and is up to 75\% in the $D^{*0}$ formation.
Simple phenomenological treatment of this process consists in supplementing the fragmentation
function with a $\delta$-term.

Author would like to thank Prof. Likhoded A.K. for useful discussions.
The work was financially supported by Russian Foundation for Basic
Research (grant \#10-02-00061a), non-commercial
foundation \textquotedblleft{}Dynasty\textquotedblright{} and the
grants of the president of Russian Federation MK-140.2009.2 and MK-406.2010.2.


\newpage

\clearpage\newpage

\end{document}